\documentclass[12pt,a4paper]{article}
\usepackage{feynmp,amstex,amssymb}

\makeatletter
\newif\if@preliminary
\@preliminaryfalse
\def\preliminary{\@preliminarytrue}
\def\preprintno#1{\def\@preprintno{#1}}
\def\address#1{\def\@address{#1}}
\def\email#1#2{\thanks{\tt #1@{}#2}}
\def\abstract#1{\def\@abstract{#1}}
\renewcommand\abstractname{ABSTRACT}
\newlength\preprintnoskip
\newlength\abstractwidth
\@titlepagetrue
\renewcommand\maketitle{\begin{titlepage}%
  \let\footnotesize\small
  \hfill\parbox{\preprintnoskip}{%
  \begin{flushright}\@preprintno\end{flushright}}\hspace*{1cm}
  \vskip 60\p@
  \begin{center}%
    {\Large\bf\boldmath \@title \par}\vskip 1cm%
    {\sc\@author \par}\vskip 3mm%
    {\@address \par}%
    \if@preliminary
      \vskip 2cm {\large\sf PRELIMINARY DRAFT \par \@date}%
    \fi
  \end{center}\par
  \@thanks
  \vfill
  \begin{center}%
    \parbox{\abstractwidth}{\centerline{\abstractname}%
    \vskip 3mm%
    \@abstract}
  \end{center}
  \end{titlepage}%
  \setcounter{footnote}{0}%
  \let\thanks\relax\let\maketitle\relax
  \gdef\@thanks{}\gdef\@author{}\gdef\@address{}%
  \gdef\@title{}\gdef\@abstract{}\gdef\@preprintno{}
}%
\def\@citex[#1]#2{\if@filesw\immediate\write\@auxout{\string\citation{#2}}\fi
  \def\@citea{}\@cite{\@for\@citeb:=#2\do
    {\@citea\def\@citea{,\penalty\@m}\@ifundefined
       {b@\@citeb}{{\bf ?}\@warning
       {Citation `\@citeb' on page \thepage \space undefined}}%
\hbox{\csname b@\@citeb\endcsname}}}{#1}}
\def\citerange{\@ifnextchar [{\@tempswatrue\@citexr}{\@tempswafalse\@citexr[]}}
\def\@citexr[#1]#2{\if@filesw\immediate\write\@auxout{\string\citation{#2}}\fi
  \def\@citea{}\@cite{\@for\@citeb:=#2\do
    {\@citea\def\@citea{--\penalty\@m}\@ifundefined
       {b@\@citeb}{{\bf ?}\@warning
       {Citation `\@citeb' on page \thepage \space undefined}}%
\hbox{\csname b@\@citeb\endcsname}}}{#1}}
\long\def\@makecaption#1#2{%
  \vskip\abovecaptionskip
  \sbox\@tempboxa{#1: \emph{#2}}%
  \ifdim \wd\@tempboxa >\hsize
    #1: \emph{#2}\par
  \else
    \hbox to\hsize{\hfil\box\@tempboxa\hfil}%
  \fi
  \vskip\belowcaptionskip}
\def\fmslash{\@ifnextchar[{\fmsl@sh}{\fmsl@sh[0mu]}}
\def\fmsl@sh[#1]#2{%
  \mathchoice
    {\@fmsl@sh\displaystyle{#1}{#2}}%
    {\@fmsl@sh\textstyle{#1}{#2}}%
    {\@fmsl@sh\scriptstyle{#1}{#2}}%
    {\@fmsl@sh\scriptscriptstyle{#1}{#2}}}
\def\@fmsl@sh#1#2#3{\m@th\ooalign{$\hfil#1\mkern#2/\hfil$\crcr$#1#3$}}
\makeatother

\newcommand{\ps}{{\text{\it P\!S}}}
\newcommand{\pom}{{\mathbb{P}}}
\newcommand{\odd}{{\mathbb{O}}}
\newcommand{\mn}{{\text{min}}}
\newcommand{\mx}{{\text{max}}}
\newcommand{\cm}{{\text{cm}}}
\begin{document}
\baselineskip20pt   
\preprintno{HD--THEP 97--57\\hep-ph/9712371\\[0.5\baselineskip] December 1997}
\title{%
 SINGLE PSEUDOSCALAR MESON PRODUCTION\\
 IN DIFFRACTIVE $ep$ SCATTERING\footnote{%
Work supported by German Bundesministerium f\"ur Bildung und
Forschung (BMBF), Contract Nr.~05~6HD~91~P(0)}
}
\author{%
 W.~Kilian\email{W.Kilian}{thphys.uni-heidelberg.de}
 and O.~Nachtmann\email{O.Nachtmann}{thphys.uni-heidelberg.de}
}
\address{%
 Institut f\"ur Theoretische Physik, Universit\"at Heidelberg,
 Philosophenweg 16\\
 D--69120 Heidelberg, Germany
}
\abstract{%
Exclusive pseudoscalar meson production in $ep$ scattering at high
energies is a direct probe for a possible ``odderon'' exchange in soft
hadronic processes.  Using a simple phenomenological ansatz for the
odderon, we demonstrate how it can be separated from the contribution
due to photon-photon fusion, and the relevant parameters be measured.
Total cross sections and differential distributions are presented for
$\pi^0,\eta,\eta'$, and $\eta_c$ production.  Results are given from
both a full calculation and one using the equivalent photon
approximation.  The accuracy of the latter is discussed.
}
\maketitle

\begin{fmffile}{epgraphs}
\section{Introduction}
Although QCD is well established as the theoretical framework of
hadronic phenomena, it has remained a great challenge to derive
results for \emph{soft} hadronic interactions from first principles,
\emph{i.e.}, starting from the Lagrangian of QCD.  In particular, one
would like to understand high-energy diffractive reactions.
Pioneering work in this direction using perturbation theory can be
found in~\citerange{LN,BFKL}.  A more general framework was
developed in~\citerange{LNm,NmS}, where both nonperturbative and
perturbative effects can be treated.  In this way, a description of
high-energy diffractive reactions in terms of the vacuum parameters of
QCD and of hadron extension parameters was achieved, which gives very
satisfactory agreement with experimental results~\cite{DFK}.

On the other hand, high-energy reactions can be described by a
Regge-pole model (for reviews, cf.\ \cite{Col}).  Its application to
diffractive reactions is very successful (\cite{DL}, cf.\
also~\cite{PDG}).  For simplicity, in this paper we will use
Regge-pole parameterizations for the hadronic amplitudes occuring in
our calculations.

Consider, for instance, elastic scattering of two hadrons~$h_{1,2}$
\begin{equation}\label{hh}
  h_1(p_1) + h_2(p_2) \to h_1(p_3) + h_2(p_4),
\end{equation}
and let $s,t,u$ be the usual Mandelstam variables
\begin{equation}
  s=(p_1+p_2)^2, \qquad
  t=(p_1-p_3)^2, \qquad
  u=(p_1-p_4)^2.
\end{equation}
The Regge-pole ansatz for the $T$-matrix element of
reaction~(\ref{hh}) reads as follows:
\begin{equation}\label{pom}
  T(s,t,u)
  = \sum_{i} c_i(t)\,(s/s_0)^{\alpha_i(t)-1}.
\end{equation}
Here the individual terms correspond to the Regge poles which can
be exchanged in the reaction~(\ref{hh}), and $\alpha_i(t)$ are their
trajectories which turn out to be linear to a good approximation:
\begin{equation}
  \alpha_i(t) = \alpha_i(0) + \alpha'_i t.
\end{equation}
While the parameters $\alpha_i(0)$ and $\alpha'_i$ which govern the $s$
and part of the $t$ dependence are observed to be universal, the
coupling parameters $c_i$ containing the spin and signature factors
and Regge residues depend on the hadrons participating in the
particular process considered.  The scale factor needed for
dimensional reasons is denoted by~$s_0$.  Each Regge pole is
associated with a family of hadrons exchanged~\cite{Col,VH}.

The Regge pole corresponding to the leading term in~(\ref{pom}) for
$s\to\infty$ is called the \emph{pomeron}.  Typical values for its
parameters are~\cite{DL}
\begin{equation}
  \alpha_\pom(0)=1.08, \qquad \alpha'_\pom= 0.25\;{\rm GeV}^{-2}.
\end{equation}
The pomeron has vacuum quantum numbers, in particular charge
conjugation $C=+1$.  The simplest description of such an interaction
in perturbative QCD is by two-gluon exchange~\cite{LN}.  Thus, one
expects the pomeron to be associated to a family of glueball states,
the lowest-lying one of these having quantum numbers $J^{PC}=2^{++}$.
Lattice calculations~\cite{MP} support a mass for this state around
$2\;{\rm GeV}$ which would fit nicely
onto the pomeron trajectory~\cite{Lan}.  The experimental
situation concerning $2^{++}$ glueball states is summarized
in~\cite{PDG}.  Detailed theoretical investigations of the pomeron in
perturbative QCD can be found in~\cite{Lip,FR}.

A natural question is whether there exist effects in high-energy
hadron-hadron scattering where the $s$-dependence is similar to the
one induced by the pomeron, but which are connected with $C=-1$
exchange.  In the framework of the Regge-pole model, the corresponding
object has been called the \emph{odderon}~\citerange{JLN,CDGJP}.
Various possibilities were discussed for it, ranging from a moving
pole, similar to the pomeron case, to exotic possibilities such as two
complex poles~\cite{BGN}.  In perturbative QCD, an odderon arises in
diagrams where three or more gluons are exchanged.  Indeed, both in
perturbative~\citerange{Lip90,JW} and in
nonperturbative~\cite{Nm,DLo} calculations one finds no reason for the
odderon contribution to be particularly small in quark-quark
scattering.  [However, since free quarks do not exist, this process
cannot be studied by itself in experiments.]  On the other hand, no
odderon has so far been observed in hadron-hadron elastic
scattering~(\ref{hh}) for $s\to\infty$, $|t|$ small.  Possible
resolution of this puzzle have been proposed in~\cite{Gin,DR}.

Thus, experimental searches for the odderon are clearly worthwhile.
Evidence for either the presence or absence of such effects will give
important clues on the structure of diffractive interactions in QCD.

In~\cite{SMN} it has been pointed out that exclusive pseudoscalar
meson production in $e^\pm p$ collisions at high energies (for HERA:
$\sqrt{s}=300.6\;{\rm GeV}$) is a direct probe for the odderon
(Fig.\ref{fig:process}):
\begin{equation}\label{process}
  e^\pm p \to e^\pm p\;\ps,
\end{equation}
where $\ps$\ generically denotes a meson with the quantum numbers
$J^{PC}=0^{-+}$, in particular, $\ps = \pi^0,\eta,\eta'$, or $\eta_c$.
Since the quantum numbers exchanged in the hadronic interaction
[Fig.\ref{fig:process}(b)] are those of the photon, the
process~(\ref{process}) can proceed also via photon-photon fusion
[Fig.\ref{fig:process}(a)].  Here and in the following we always work
in leading-order perturbation theory of the electroweak interactions.
Adopting standard parameterizations for the proton and meson form
factors, the diagram in Fig.\ref{fig:process}(a) can easily be
calculated.  Note that the exchange of an object with vacuum quantum
numbers, \emph{i.e.}, of the pomeron, is forbidden in
reaction~(\ref{process}).

In the present paper we will extend the considerations in~\cite{SMN}
and study, from a purely phenomenological point of view, the effect of
an odderon interaction on the process~(\ref{process}) and the
possibilities to extract detailed information about its properties.

\newcommand{\grapha}{%
\unitlength1mm%
\begin{fmfgraph*}(50,40)
  \fmfsurroundn{e}{16}
  \fmf{fermion}{e3,v1}\fmf{fermion,tens=2}{v1,e7}
  \fmf{photon}{v1,v3}
  \fmf{double}{v3,e1}
  \fmf{photon}{v2,v3}
  \fmf{heavy}{v2,e15}\fmf{heavy,tens=2}{e11,v2}
  \fmfblob{5mm}{v2,v3}
  \fmflabel{$e^\pm(p_1)$}{e7}
  \fmflabel{$e^\pm(p_1')$}{e3}
  \fmfv{l=$\gamma(q_1)$,l.d=5mm,l.a=-114}{v1}
  \fmflabel{$\ps(k)$}{e1}
  \fmfv{l=$\gamma(q_2)$,l.d=5mm,l.a=114}{v2}
  \fmflabel{$p(p_2)$}{e11}
  \fmflabel{$p(p_2')$}{e15}
  \fmfv{l={ $s_1$},l.d=0}{e2}
  \fmfv{l={ $s_2=W^2$},l.d=0}{e16}
  \fmfv{l={ $s$},l.d=0}{e9}
  \fmfv{l={ $t_1=-Q^2$},l.d=0}{e5}
  \fmfv{l={ $t_2$},l.d=0}{e13}
\end{fmfgraph*}%
}

\newcommand{\graphb}{%
\unitlength1mm%
\begin{fmfgraph*}(50,40)
  \fmfsurroundn{e}{16}
  \fmf{fermion}{e3,v1}\fmf{fermion,tens=2}{v1,e7}
  \fmf{photon}{v1,v3}
  \fmf{double}{v3,e1}
  \fmf{zigzag}{v2,v3}
  \fmf{heavy}{v2,e15}\fmf{heavy,tens=2}{e11,v2}
  \fmfblob{5mm}{v2,v3}
  \fmflabel{$e^\pm$}{e7}
  \fmflabel{$e^\pm$}{e3}
  \fmfv{l=$\gamma$,l.d=5mm,l.a=-114}{v1}
  \fmflabel{$\ps$}{e1}
  \fmfv{l=$\odd$,l.d=5mm,l.a=114}{v2}
  \fmflabel{$p$}{e11}
  \fmflabel{$p$}{e15}
\end{fmfgraph*}%
}

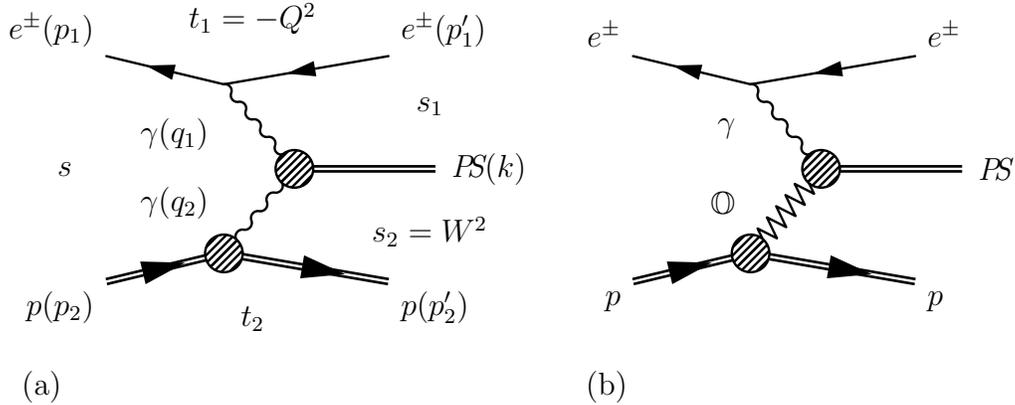
\begin{figure}
\begin{center}\unitlength1mm
\begin{picture}(140,55)
  \put(0,0){(a)}
  \put(5,10){\grapha}
  \put(75,0){(b)}
  \put(75,10){\graphb}
\end{picture}
\caption{Feynman diagrams for pseudoscalar meson production in $ep$
scattering at high energies with photon (a) and odderon (b) exchange.}
\label{fig:process}
\end{center}
\end{figure}

The kinematical variables are as indicated in
Fig.\ref{fig:process}(a).  In addition, in accordance with the usual
notation for deep-inelastic scattering processes, we introduce the
fractional energy loss of the electron 
\begin{equation}\label{HERA}
  y = \frac{p_2\cdot(p_1-p_1')}{p_2\cdot p_1},
\end{equation}
where $0\leq y\leq 1$.  

If the outgoing proton is not observed, the signal consists
of the outgoing electron and of the decay products of the $\ps$\
particle, \emph{e.g.}, a $\gamma\gamma$ pair.  The numbers and
distributions presented in the following sections refer either to the
complete phase space or to two particular sets of cuts appropriate for
the HERA environment~\cite{ST} defined and denoted as follows:

\begin{alignat}{3}
  0.3&<y<0.7,\qquad  &0&<Q^2<0.01\;{\rm GeV}^2, \qquad 
  &&\emph{Photoproduction (PP)},
  \label{cuts-pp}
  \\
  0.3&<y<0.7,\qquad  &1\;{\rm GeV}^2&<Q^2, 
  &&\emph{Deep inelastic scattering (DIS)},
  \label{cuts-dis}
\end{alignat}
where $Q^2\equiv -q_1^2$.

\section{Meson form factor parameterization}
The coupling of a pseudoscalar meson $\ps=\pi^0,\eta,\eta',\eta_c$ to
two photons has the form
\begin{equation}\label{coupling}
  iu\epsilon_{\mu\nu\rho\sigma}q_1^\rho q_2^\sigma
       \,T(q_1^2,q_2^2),
\end{equation}
where, by definition, $T(0,0)=1$.  The dimensionful coupling constant $u$
is related to the triangle anomaly.  In terms of the meson decay
constant $f_\ps$, it is given by 
\begin{equation}\label{u-fps}
  u = \alpha/(\pi f_\ps).
\end{equation}
We choose a convention where $f_\pi=93\;{\rm MeV}$, $\alpha\equiv
e^2/4\pi$ is the fine-structure constant, and $e$ the proton charge.
From~(\ref{coupling},\ref{u-fps}) we find for the partial
width~$\Gamma_{\gamma\gamma}$ for $\ps\to\gamma\gamma$:
\begin{equation}
  \Gamma_{\gamma\gamma} = \frac{u^2}{64\pi} m_\ps^3
\end{equation}
For our numerical results we use thus
\begin{equation}\label{u}
  u = \sqrt{64\pi\Gamma_{\gamma\gamma}/m_\ps^3},
\end{equation}
with $\Gamma_{\gamma\gamma}$ taken from experiment (cf.\
Tab.\ref{tab:mesons}).

\begin{table}[bt]
\begin{center}
$
\begin{array}{|l||c|c|c|c|}
\hline
 & \pi^0 & \eta & \eta' & \eta_c \\
\hline
 m\;[{\rm MeV}]
   & 134.9764(5) & 547.45(19) & 957.77(14) & 2979.8(2.1) \\
 \Gamma_{\rm tot}\;[{\rm MeV}]
   & 7.8(6)\times 10^{-6} & 1.18(11)\times 10^{-3} & 0.201(16) & 13.2(3.8)\\
 {\rm BR}(\gamma\gamma)\;[\%]
   & 98.798(32) & 39.25(31) & 2.12(13) & 0.30(12) \\
\hline
 u\;[{\rm GeV}^{-1}]
   & 0.025(1) & 0.024(1) & 0.031(1) & 0.0075(15) \\
\hline
\end{array}
$
\caption{Pseudoscalar meson data, from~\protect\cite{PDG}.}
\label{tab:mesons}
\end{center}
\end{table}

The form factor $T(q_1^2,q_2^2)$ is known as the \emph{transition form
factor} of the pseudoscalar meson.  It can be measured in pseudoscalar
production in $e^+e^-$ scattering, $e^+e^-\to e^+e^-\ps$.  The diagram
for this reaction is as in Fig.\ref{fig:process}(a), but with the
proton replaced by an electron line.  It is found that the formula
\begin{equation}\label{BL-interpol}
  T(q_1^2,0) = \frac{1}{1-q_1^2/8\pi^2 f_\ps^2}
\end{equation}
introduced by Brodsky and Lepage~\cite{BL81}, which is confirmed by
constituent quark model~\cite{AMN} and QCD sum-rule
calculations~\cite{RR}, fits the data for $q_1^2\neq 0$, $q_2^2 = 0$
reasonably well~\cite{Exp}.  Within the experimental errors, the
formula~(\ref{BL-interpol}) coincides in this kinematic region with
the double pole form suggested by vector meson dominance
\begin{equation}\label{2pole}
  T(q_1^2,q_2^2) = \frac{1}{(1-q_1^2/\Lambda^2)(1-q_2^2/\Lambda^2)},
\end{equation}
where for the light mesons, $\Lambda$ is given by the $\rho$ (or
$\omega$) meson mass.  For $\eta_c$ production, one expects a slower
decrease of the form factor with $q_i^2$~\cite{FK}; one should then
insert the $J/\psi$ mass for $\Lambda$ in~(\ref{2pole}).

\begin{figure}
\begin{center}
\includegraphics{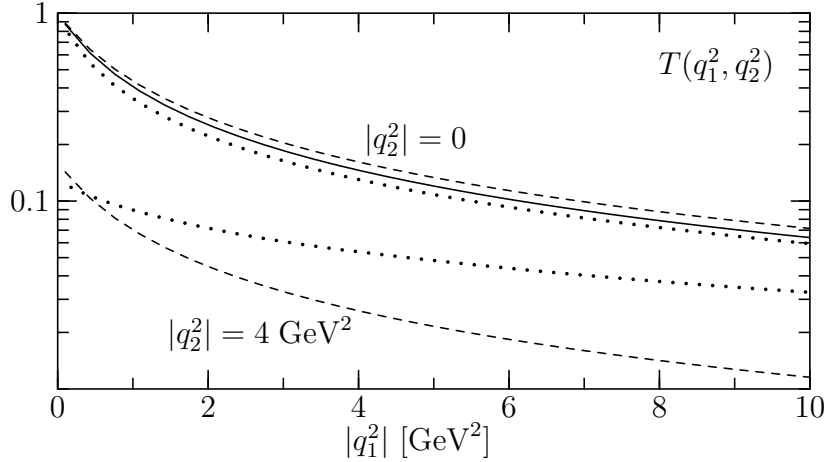}
\end{center}
\caption{Various parameterizations for the pion transition form factor
$T(q_1^2,q_2^2)$.  The solid curve represents the Brodsky-Lepage
formula~(\ref{BL-interpol}).  The dashed and dotted curves correspond
to the double pole form~(\ref{2pole}) and the interpolation
formula~(\ref{Tnew}), respectively.}
\label{fig:T}
\end{figure}

Precise data are available only for $T(q^2,0)$, \emph{i.e.}, with one
photon nearly on-shell.  For both photons far off-shell, perturbative
QCD predicts~\cite{KWZ}
\begin{equation}\label{PQCD}
  T(q_1^2, q_2^2) = -\frac{8\pi^2}{3} f_\ps^2
	\int_0^1 dx
	\frac{\varphi(x,\bar x)}{x q_1^2 + \bar x q_2^2},
\end{equation}
with $\bar x\equiv 1-x$.  The amplitude $\varphi(x,\bar x)$ is
normalized ($\int dx\,\varphi(x,\bar x)=1$) and has the asymptotic
form
\begin{equation}
  \varphi(x,\bar x) = 6x\bar x.
\end{equation}
The form~(\ref{2pole}) is not compatible with~(\ref{PQCD}).  One might
worry whether this discrepancy has an impact on the quantitative
predictions for pseudoscalar meson productions that will be presented
below.  For comparison, we introduce a formula which interpolates
between the asymptotic limit~(\ref{PQCD}) and the on-shell limit $T=1$
for both $q^2$ nonvanishing:
\begin{equation}\label{Tnew}
\begin{split}
  T(q_1^2,q_2^2) &= -\frac{8\pi^2}{3} f_\ps^2
	\int_0^1 dx
	\frac{\varphi(x,\bar x)}{x q_1^2 + \bar x q_2^2 
		- 8\pi^2 f_\ps^2/3}
	\\[.5\baselineskip]
	&= 8\pi^2 f_\ps^2
	\left(-\frac{\hat q_1^2 +\hat q_2^2}{(\hat q_1^2-\hat q_2^2)^2}
	+ \frac{2\hat q_1^2\hat q_2^2}{(\hat q_1^2-\hat q_2^2)^3}
	\ln\frac{\hat q_1^2}{\hat q_2^2}\right),
\end{split}
\end{equation}
where $\hat q_i^2 \equiv q_i^2 - 8\pi^2 f_\ps^2/3$.  This form agrees
with~(\ref{BL-interpol}) for $q_1^2\neq 0$, $q_2^2=0$, and
with~(\ref{PQCD}) for $q_{1,2}^2\neq 0$, up to corrections which scale
like $\log|q_i^2|/q_i^4$ in the limit $q_i^2\to -\infty$.  In
Fig.\ref{fig:T} we compare the shape of the above form factor
expressions for two fixed values of $q_2^2$.  At $q_2^2=0$ all three
curves are close to each other.  By contrast, at $q_2^2=4\;{\rm
GeV}^2$, the interpolation curve~(\ref{Tnew}) lies considerably higher
than the naive double pole ansatz~(\ref{2pole}), which we use as
standard parameterization in the following.  However, we find that if
the alternative form factor~(\ref{Tnew}) is inserted, the shift in the
cross section values and distributions discussed below is numerically
below 1\% in all cases.

\section{Results for the two-photon process}
Once the form factors are given, the photon-photon amplitude
[Fig.\ref{fig:process}(a)] is determined completely.  The spin-averaged
squared matrix element has the form~\cite{BGMS}
\begin{equation}\label{ME}
  \frac14|M|^2 =
  2\Delta\frac{u^2|T(t_1,t_2)|^2}{t_1t_2}\rho^{++}_e\rho^{++}_p,
\end{equation}
where
\begin{equation}
  \Delta
  = \frac14(m_\ps^2-t_1-t_2)^2 - t_1 t_2
\end{equation}
is the phase space function for the subprocess
$\gamma^*\gamma^*\to\ps$.  We have assumed that the azimuthal angles
of the outgoing particles are not observed.  The probabilities for
photon emission off the initial electron/positron and proton,
are given by $\rho^{++}_e$ and $\rho^{++}_p$, respectively, with
\begin{eqnarray}
  \rho_e^{++} &=&  1  + 
	2\frac{m_e^2}{q_1^2} + \frac{1}{2\Delta}
	\left[4(p_1q_2)(p_1'q_2) + q_1^2 q_2^2 \right]
  \\
  \rho_p^{++} &=&  C(t_2)  + D(t_2)
	\left(2\frac{m_p^2}{q_2^2} + \frac{1}{2\Delta}
	\left[4(p_2q_1)(p_2'q_1) + q_1^2 q_2^2 \right]\right)
\end{eqnarray}
The coefficient functions $C$ and $D$ are determined by the electric
and magnetic form factors of the proton:
\begin{equation}
  C(t_2) = G_M^2(t_2), \qquad
  D(t_2) = \frac{4m_p^2 G_E^2(t_2) - t_2 G_M^2(t_2)}{4m_p^2 - t_2}
\end{equation}
We use the standard dipole parameterization
\begin{equation}
  G_E(t) = \frac{1}{(1-t/m_D^2)^2}, \qquad
  G_M(t) = \mu_p\,G_E(t),
\end{equation}
with $m_D^2=0.71\;{\rm GeV}^2$ and $\mu_p=2.7928$~\cite{pFF}.

Finally, we note that longitudinal photon polarizations do not
contribute in~(\ref{ME}) due to the coupling~(\ref{coupling}).

For the calculation of two-photon processes, equivalent-photon
approximations (EPA) are commonly used.  Therefore, in the following
sections we compare two different versions of the EPA with a
numerical integration of the exact expression for the cross section,
and discuss the accuracy of the approximations in various kinematical
regions.

\subsection{Double Equivalent Photon Approximation (DEPA)}
Since the dominant contribution to the $\gamma\gamma$-fusion process
arises from the region in phase space where both photon virtualities
are small, \emph{i.e.}, $|t_1|,|t_2|\ll m^2_\ps, m^2_\rho$, an
estimate for the total cross section may be found from the
DEPA~\cite{BGMS}.  Introducing the fractional energy losses
\begin{equation}
  x_i = \frac{E_i-E_i'}{E_i},\qquad \bar x_i\equiv 1-x_i
\end{equation}
of the incoming $e^\pm$ ($i=1$) and proton ($i=2$), the total cross
section is given by
\begin{equation}\label{sigma-DEPA}
  \sigma(ep\to ep+\ps) =
  \int_0^1 \frac{dx_1}{x_1}\,\frac{dx_2}{x_2}\,
	n_1(x_1)\,n_2(x_2)\,
	\frac{\pi u^2 m_\ps^2}{8s}
	\delta(x_1x_2-m_\ps^2/s)
\end{equation}
with the photon fluxes
\begin{equation}
  n_i(x) = \frac{\alpha}{\pi}\left[
	\left(1-x+\mu_i^2\frac{x^2}{2}\right)
	\ln\frac{t_{i,\,\mn}}{t_{i,\,\mx}}
	- (1-x)\left(1 - \frac{t_{i,\,\mn}}{t_{i,\,\mx}}\right)
	\right],
\end{equation}
where $\mu_i$ is the magnetic moment of the incoming $e^\pm$ resp.\
proton.  The photon virtualities $t_i$ have been integrated over the
range
\begin{equation}
  t_{i,\,\mn} \equiv-\Lambda^2 = -\min(m_\ps^2,m_\rho^2,q_0^2), \qquad
  t_{i,\,\mx} = -\frac{x_i^2}{1-x_i}m_i^2.
\end{equation}
where $m_{1,2}$ is given by $m_e$ and $m_p$, respectively.
Within the validity of the DEPA, it is legitimate to replace the
detailed $t_i$ dependence of the meson and proton form factors by an
appropriate sharp cutoff~$\Lambda^2$.

The delta function in~(\ref{sigma-DEPA}) restricts the integration
range for the independent $x$ variable between
$((m_p+m_\ps)^2-m_p^2)/s$ and~$1$.  The error of the approximation is
of the order
\begin{equation}
  \Delta\sigma/\sigma \sim \frac{1}{\ln |t_\mx|/\Lambda^2}.
\end{equation}
which depends on $x_1$ and $x_2$.

Within the validity range of the DEPA, the following simplifications
apply (cf.\ Fig.\ref{fig:process}):
\begin{equation}
  y = x_1, \qquad  W^2 = y s, \qquad  s_1 = \frac{m_\ps^2}{y},
	\qquad Q^2=0.
\end{equation}
Furthermore, the rapidity of the produced pseudoscalar has the value
\begin{equation}\label{rap}
  \eta = \eta_\cm - \frac12\log\frac{x_1}{x_2} 
  = \eta_\cm - \log\frac{\sqrt{s}}{m_\ps} - \log y,
\end{equation}
where $\eta_\cm$ is the rapidity of the c.m.\ system of the
incoming particles. [At HERA, $\eta_\cm=1.69$, where we choose the
usual convention that the initial proton momentum points in positive
$z$-direction.]

\subsection{Single Equivalent Photon Approximation (EPA)}
This approximation can be used in the photoproduction region where
$-t_1=Q^2<0.01\;{\rm GeV}^2$ is small, but $|t_2|$ can become large.
Then the DEPA is no longer applicable.  However, the photon
$\gamma(q_1)$ emitted by the initial electron/positron is still
essentially on-shell and may be treated within the EPA.  The
corresponding approximation to the total cross section reads
\begin{equation}\label{EPA}
  \sigma(s) = \int_{y_\mn}^{y_\mx} \frac{dy}{y} n(y)\,
	\int_{t_\mn}^{t_\mx} dt_2\,
	\frac{d\hat\sigma}{dt_2}
	\qquad\mbox{with}\qquad
	s_2=ys + (1-y)m_p^2,
\end{equation}
where
\begin{equation}
\begin{split}
  \frac{d\hat\sigma}{dt_2}
  &=
	\frac{e^2 u^2 |T(t_2,0)|^2}{64\pi(s_2-m_p^2)^2 t_2}
	\nonumber\\
  &\quad\times\big\{
	|F_1(t_2)|^2\big[
	-2 t_2^{-1} m_p^2 m_\ps^4
	-2m_p^4 + 2 m_p^2 m_\ps^2 - m_\ps^4
		+ 4m_p^2 s_2 + 2 m_\ps^2 s_2 - 2 s_2^2
	\\
  &\quad\phantom{\times\big\{|F_1(t_2)|^2\big[}
	+ t_2(2m_\ps^2 - 2s_2) - t_2^2\big]
	\\
  &\quad\phantom{\times\big\{}
	+ \frac{1}{2m_p^2}|F_2(t_2)|^2\big[
	-m_p^2 m_\ps^4
	+ t_2(m_p^4 + 3 m_p^2 m_\ps^2 - 2 m_p^2 s_2
		- m_\ps^2 s_2 + s_2^2)
	\\
  &\quad\phantom{\times\big\{+ \frac{1}{2m_p^2}|F_2(t_2)|^2\big[}
	+ t_2^2(-2m_p^2 + s_2)\big]\big\}.
\end{split}
\end{equation}
Here $F_{1,2}(t_2)$ are the Dirac and Pauli form factors, which are
related to $G_E,G_M$ by
\begin{equation}
  G_E(t_2)\equiv F_1(t_2) + t_2\frac{1}{4m_p^2}F_2(t_2), \qquad
  G_M(t_2)\equiv F_1(t_2) + F_2(t_2).
\end{equation}
The kinematic limits for the integrations in~(\ref{EPA}) are given by
\begin{equation}
  y_\mn = \frac{(2m_p+m_\ps)m_\ps}{s-m_p^2},
	\qquad
  y_\mx = 1,
\end{equation}
and
\begin{equation}
  t_{\rm min,\;max} =
	\frac{m_\ps^4}{4s_2} -
	\left[\frac{s_2-m_p^2}{2\sqrt{s_2}}
	\pm \sqrt{\frac{(s_2+m_\ps^2-m_p^2)^2}{4s_2} - m_\ps^2}\;
	\right]^2.
\end{equation}
The above formulae can easily be implemented in a Monte Carlo event
generator, which we have used to generate the distributions in the PP
region in the following sections.  The error of the approximation
depends on $y$; for $y$ values of order unity, it is usually estimated
as~\cite{BGMS}
\begin{equation}\label{EPA-error}
  \Delta\sigma/\sigma \sim \frac{Q^2_\mx}{\Lambda^2}\,
	\frac{1}{\ln Q^2_\mx/m_e^2},
\end{equation}
where $Q^2_\mx=0.01\;{\rm GeV}^2$ for the PP
cuts~(\ref{cuts-pp}), and $\Lambda$ is the effective cutoff which
enters the $\ps$\ form factor.

\subsection{Full calculation}  
In order to have a reliable result in all regions of phase space, an
exact integration of the squared matrix element~(\ref{ME}) is
necessary.  This is nontrivial in practice since at the edge of phase
space large cancellations occur in~(\ref{ME}).  To obtain numerically
stable results, we used the \texttt{CompHEP}~\cite{CompHEP}
kinematical module with quadruple precision numerics for the
integration and event generation\footnote{Recently, a Monte Carlo
generator which includes the full kinematical dependence has been
developed for the analogous $\gamma\gamma$ processes in $e^+e^-$
collisions~\cite{Sch}.}.

In Tab.\ref{tab:tot} we display the total cross sections for
$\pi^0,\eta,\eta'$, and $\eta_c$ production at HERA as well as the
cross sections in the PP and DIS regions, using the three methods of
calculation introduced above.  Note that there is an overall error in
the cross sections coming from the experimental uncertainty in the
mesonic $\gamma\gamma$ width (cf.\ Tab.\ref{tab:mesons}) which amounts
to $10\%$ for $\pi^0$, $\eta$, and $\eta'$, and is as large as $50\%$
for the $\eta_c$ meson.

Whereas the DEPA is an approximation at the $20\%$ level, the EPA
gives quite accurate results, in particular in the PP region.  It is
interesting, however, that for $\pi^0$ photoproduction the standard
formula~(\ref{EPA-error}) underestimates the actual error of the EPA
by more than a factor~10.  This is due to the fact that the
photon-proton scattering amplitude is singular at $t_2=0$, which
introduces a logarithmic dependence on the lower kinematical limit of
$|t_2|$.  For large $s_2$ this limit simplifies to
\begin{equation}
  t_\mx \approx - \frac{m_p^2(m_\pi^2+Q^2)^2}{s_2^2}.
\end{equation}
The EPA neglects the nonzero value of $Q^2$.  As a result, the EPA
estimate has an additional error of magnitude
\begin{equation}
  \Delta\sigma/\sigma \sim
  \frac{Q_\mx^2}{m_\pi^2}\,\frac{1}{\ln s/m_\pi^2}
  \approx 5\%.
\end{equation}
In the DIS region neither the DEPA nor the EPA are reliable, and we
just quote the result from the full calculation.

\begin{table}
\begin{center}
$
\begin{array}{|ll||c|c|c|c|}
\hline
 && \pi^0 & \eta & \eta' & \eta_c \\
\hline
  \mbox{tot} & \mbox{DEPA} 
  & 1500 & 1200 & 1700 & 65 \\
             & \mbox{EPA}  
  & 1803 & 1011 & 1320 & 51.0 \\
             & \mbox{full} 
  & 1801 & 983 & 1276 & 50.3 \\
\hline
  \mbox{PP} & \mbox{DEPA} 
  & 84 & 62 & 93 & 4.2 \\
             & \mbox{EPA}
  & 78.1 & 56.4 & 83.6 & 3.83 \\
             & \mbox{full} 
  & 74.7 & 56.3 & 83.5 & 3.84 \\
\hline
  \mbox{DIS} & \mbox{full} 
  & 0.46 & 0.41 & 0.65 & 0.49 \\
\hline
\end{array}
$
\caption{Total cross sections in pb for pseudoscalar meson production
at HERA.  The kinematical regions are defined in~(\ref{cuts-pp})
and~(\ref{cuts-dis}), and the calculational methods are described in
the text.  The statistical errors of the Monte Carlo integration are
at the permille level.}
\label{tab:tot}
\end{center}
\end{table}

\section{Odderon exchange}
\label{sec:odderon}

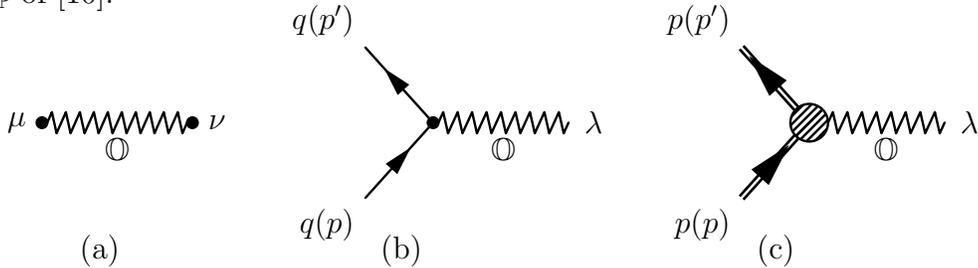
\begin{figure}[bth]
\begin{center}\unitlength1mm
\begin{fmfgraph*}(130,25)
\put(10,-3){(a)}
\begin{fmfsubgraph}(5mm,5mm)(20mm,20mm)
  \fmfleft{i}\fmfright{o}
  \fmf{zigzag, lab=$\odd$}{i,o}
  \fmfv{lab=$\mu$,l.a=180}{i} 
  \fmfv{lab=$\nu$,l.a=0}{o}
  \fmfdot{i,o}
\end{fmfsubgraph}
\fmffreeze
\put(50,-3){(b)}
\begin{fmfsubgraph}(45mm,5mm)(30mm,20mm)
  \fmfleft{i1,i2}\fmfright{o1}
  \fmf{fermion}{i1,v1,i2}  \fmf{zigzag, lab=$\odd$}{v1,o1}
  \fmfv{lab=$q(p)$,l.a=-135}{i1}  
  \fmfv{lab=$q(p')$,l.a=135}{i2}  
  \fmfv{lab=$\lambda$,l.a=0}{o1}
  \fmfdot{v1}
\end{fmfsubgraph}
\fmffreeze
\put(100,-3){(c)}
\begin{fmfsubgraph}(95mm,5mm)(30mm,20mm)
  \fmfleft{i3,i4}\fmfright{o2}
  \fmf{heavy}{i3,v2,i4}  \fmf{zigzag, lab=$\odd$}{v2,o2}
  \fmfv{lab=$p(p)$,l.a=-135}{i3}  
  \fmfv{lab=$p(p')$,l.a=135}{i4}  
  \fmfv{lab=$\lambda$,l.a=0}{o2}
  \fmfblob{5mm}{v2}
\end{fmfsubgraph}
\end{fmfgraph*}
\end{center}
\caption{Diagrammatic representations of the odderon
propagator~(\ref{odd-prop}) and its couplings to
quarks~(\ref{odd-q-coupl}) and protons~(\ref{odd-p-coupl}).}
\label{fig:Odderon-rules}
\end{figure}

Introducing now a possible odderon exchange contribution
to~(\ref{process}) [cf.\ Fig.\ref{fig:process}(b)], we make an ansatz
similar to the one made in~\cite{DL} for the pomeron.  Thus, we assume
the effective odderon ``propagator'' (Fig.\ref{fig:Odderon-rules}) to
be given by
\begin{equation}\label{odd-prop}
  (-i)\eta_\odd\,(-is/s_0)^{\alpha_\odd(t)-1}g^{\mu\nu}
\end{equation}
where $\alpha_\odd(t)$ is the odderon trajectory which we assume to be
linear
\begin{equation}\label{odd-traj}
  \alpha_\odd(t) = \alpha_\odd(0) + \alpha'_\odd t
\end{equation}
In~(\ref{odd-prop}), $\eta_\odd=\pm 1$ determines the phase of the
odderon amplitudes, which is not known \emph{a priori}.  The odderon
couplings are given as follows:

For the quark-odderon coupling (Fig.\ref{fig:Odderon-rules}b) we
set
\begin{equation}
\label{odd-q-coupl}
  -i\beta_\odd\gamma^\lambda
\end{equation}
and for the proton-odderon coupling  (Fig.\ref{fig:Odderon-rules}c)
\begin{equation}
\label{odd-p-coupl}
  -i\,3\beta_\odd
  \left[ F_1^{(0)}(q^2)\,\gamma^\lambda
	+ \frac{i}{2m_p}F_2^{(0)}(q^2)\,\sigma^{\lambda\nu}q_\nu
  \right] \qquad
  \mbox{with $q=p'-p$}.
\end{equation}
Here
\begin{equation}\label{pn-ff}
  F_i^{(0)}(q^2) = F_i^p(q^2) + F_i^n(q^2), \qquad i=1,2
\end{equation}
are the isoscalar nucleon form factors, and $\beta_\odd$ is the
analogue of the quark-pomeron coupling constant $\beta_\pom$
of~\cite{DL}.  

In the following we will neglect the neutron form factors
in~(\ref{pn-ff}) and set $F_i^{(0)}(q^2)=F_i^p(q^2)$ for simplicity.
This is a good approximation for the Dirac form factor
$F_1^{(0)}(q^2)$ since $|F_1^n(q^2)|\ll |F_1^p(q^2)|$, and the
contribution of the Pauli form factor $F_2^{(0)}(q^2)$ is small in the
region of small $|q^2|$ where most of the cross section in the
processes considered here comes from.

The difference of the ratios~$\rho$ of the real and imaginary parts of
the forward $\bar pp$ and $pp$ scattering amplitudes is then given by
\begin{equation}\label{op-ratio}
\begin{split}
  \rho^{\bar pp}(s) - \rho^{pp}(s)
  &= -2\eta_\odd\left(\frac{\beta_\odd}{\beta_\pom}\right)^2
	\left(\frac{s}{s_0}\right)^{\alpha_\odd(0)-\alpha_\pom(0)}
	\frac{\cos\left[\frac{\pi}{2}(\alpha_\odd(0)-1)\right]}
	     {\cos\left[\frac{\pi}{2}(\alpha_\pom(0)-1)\right]} \\
  &\quad\times
	\left[1 + {\cal O}(\beta_\odd^2/\beta_\pom^2,\,
			\alpha_\pom(0)-1)\right].
\end{split}
\end{equation}
where we assume the model of~\cite{DL} for pomeron exchange.

At energies $\sqrt{s}\gtrsim 100\;{\rm GeV}$ where non-leading Regge
pole contributions should be negligible, we have results only for
$\rho^{\bar pp}$ from $\bar pp$ collision experiments~\cite{UA4}.  One
has to resort to dispersion theory calculations (cf., \emph{e.g.},
\cite{CDGJP}) to extract $\rho^{pp}$ from data.  We take as an
estimate
\begin{equation}
  |\rho^{\bar pp}(s)-\rho^{pp}(s)| 
  \lesssim 0.05 \qquad
  \mbox{for $\sqrt{s}>100\;{\rm GeV}$}.
\end{equation}
To translate this into information on the ratio
$\beta_\odd/\beta_\pom$ we still need to know the value of
$\alpha_\odd(0)-\alpha_\pom(0)$ in~(\ref{op-ratio}).  Assuming, for
instance, $\alpha_\odd(0)=1$, which is suggested by the
field-theoretic arguments of~\cite{Nm} and the results of~\cite{AB},
we get
\begin{equation}\label{bound}
  (\beta_\odd/\beta_\pom)^2 \lesssim 0.05.
\end{equation}
Certainly, we do \emph{not} claim this to be a bound on odderon
contributions in $pp$ and $\bar pp$ scattering.  But~(\ref{bound})
should give an idea on the strength of a possible odderon in
conventional Regge parameterizations.

\begin{figure}
\begin{center}\unitlength1mm
\begin{fmfgraph*}(110,30)
\put(10,-3){(a)}
\begin{fmfsubgraph}(5mm,5mm)(40mm,25mm)
  \fmfleft{i1,i2} \fmfright{o1}
  \fmf{boson}{i1,v1} \fmf{boson}{i2,v2}  \fmf{double}{v3,o1}
  \fmf{fermion, tens=1/3}{v1,v2,v3,v1}
  \fmfv{lab=$\gamma$,l.a=-135}{i1} 
  \fmfv{lab=$\gamma$,l.a=135}{i2}
  \fmfv{lab=$\ps$,l.a=0}{o1}
\end{fmfsubgraph}
\fmffreeze
\put(70,-3){(b)}
\begin{fmfsubgraph}(65mm,5mm)(40mm,25mm)
  \fmfleft{i3,i4} \fmfright{o2}
  \fmf{zigzag}{i3,v4} \fmf{boson}{i4,v5}  \fmf{double}{v6,o2}
  \fmf{fermion, tens=1/3}{v4,v5,v6,v4}
  \fmfv{lab=$\odd$,l.a=-135}{i3} 
  \fmfv{lab=$\gamma$,l.a=135}{i4}
  \fmfv{lab=$\ps$,l.a=0}{o2}
\end{fmfsubgraph}
\fmfdotn{v}{6}
\end{fmfgraph*}
\end{center}
\caption{Diagrams for the $\gamma\gamma\ps$ and $\gamma\odd\ps$ form
factors.  The crossed diagrams are to be added.}
\label{fig:formfactors}
\end{figure}
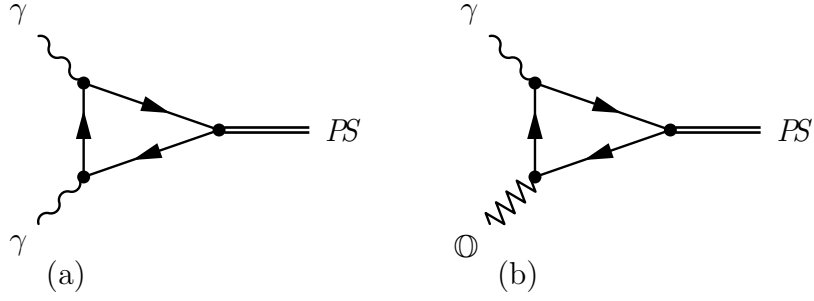

Considering the ratio of the odderon and photon couplings to the
produced meson, one naturally expects that the
$\gamma\gamma$ coupling is proportional to the
electromagnetic charge squared of the quark states within the meson,
whereas for the $\gamma\odd$ coupling only a single charge factor
enters, if the odderon is assumed to be flavor-blind.  Thus, for a
meson of quark content
\begin{equation}
  \ps \sim \sum_i a^i_\ps\, q_i\bar q_i,
\end{equation}
in a valence quark model we draw for the $\gamma\gamma\ps$ and
$\gamma\odd\ps$ form factors $T^{\gamma\gamma}$ [$\equiv T$ of
(\ref{coupling})] and $T^{\gamma\odd}$
the diagrams shown in Fig.\ref{fig:formfactors}.  Assuming all
\ps-wave function effects to be identical in the diagrams (a) and (b)
of Fig.\ref{fig:formfactors}, and denoting by $Q_i$ the quark charge
in units of the proton charge, we get
\begin{equation}
  \frac{T^{\gamma\odd}}{T^{\gamma\gamma}}
  = \frac{e\beta_\odd\sum_i a^i_\ps\,Q_i}{e^2\sum_i a^i_\ps\,Q_i^2}
  \equiv \frac{\beta_\odd}{e}\,r_\ps.
\end{equation}
Within this valence quark model, the relative coupling strengths may then
be derived by assuming that the pion is an isospin triplet, the $\eta$
and $\eta'$ are mixtures of $SU(3)$ singlet and octet with mixing
angle~$\theta$, and the $\eta_c$ is a pure $c\bar c$ state:
\begin{equation}\label{model}
  r_\pi = 3, \qquad
  r_\eta = 3\cos\theta, \qquad
  r_{\eta'} = -3\sin\theta, \qquad
  r_{\eta_c} = 3/2.
\end{equation}
Since a $SU(3)$ singlet decouples, in the limit $\theta=0$ the
odderon coupling of the $\eta'$ vanishes.  In the following analysis
we will take the value $\theta=-20^\circ$~\cite{PDG}.

With these assumptions, the amplitude for the sum of the diagrams
Fig.\ref{fig:process}(a,b) is obtained from the
$\gamma\gamma$-amplitude by a simple replacement for the term
corresponding to the propagator of the photon $\gamma(q_2)$:
\begin{equation}\label{replace}
  \frac{e^2}{t_2} \longrightarrow
  \frac{e^2}{t_2} + 3r_\ps\,\eta_\odd\,\beta_\odd^2\,
	\left(-i\frac{s_2}{s_0}\right)^{\alpha_\odd(t_2)-1}
\end{equation}
Our ansatz depends on the quark-odderon coupling $\beta_\odd$, on the
phase $\eta_\odd$, and on the three parameters
$\alpha_\odd(0),\alpha'_\odd,s_0$.  As reference values we choose
\begin{equation}\label{ref}
  \alpha_\odd(0) = 1, \qquad 
  \alpha'_\odd=0.25\;{\rm GeV}^{-2}, \qquad
  s_0 = 1\;{\rm GeV}^2, \qquad
  \eta_\odd=-1, \qquad
  \beta_\odd^2 = 0.05\,\beta_\pom^2,
\end{equation}
where $\beta_\pom=1.8\;{\rm GeV}^{-1}$ is the quark-pomeron coupling.
The abbreviation
\begin{equation}
  c_\odd = \eta_\odd\,\beta_\odd^2/\beta_\pom^2
\end{equation}
turns out to be convenient for the presentation of our numerical
results below.

\begin{figure}
\begin{center}
\includegraphics{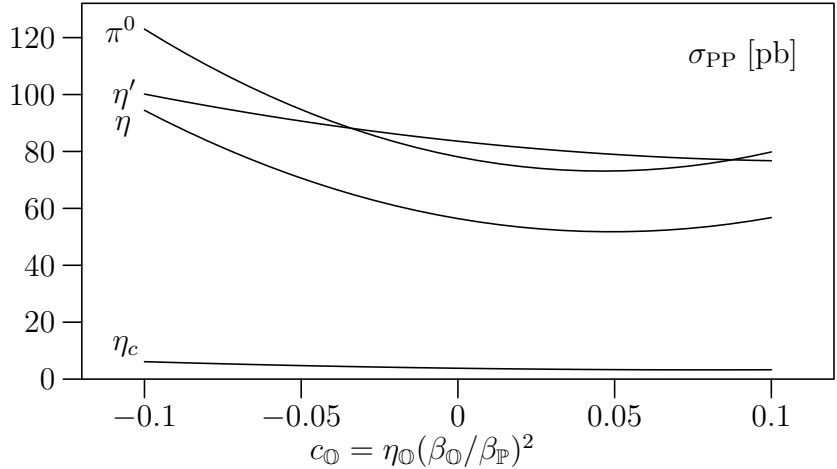}
\end{center}
\caption{Total cross section for pseudoscalar meson production in the
photoproduction region~(\ref{cuts-pp}) as a function of the odderon
coupling $c_\odd$.  The other parameters are taken at their reference
values~(\ref{ref}), and the relative coupling strengths are taken
from~(\ref{model}).}
\label{fig:model}
\end{figure}

By comparing the measured total cross sections in different channels
with the expectation from photon-photon fusion alone, the odderon
couplings can in principle be determined and the model
values~(\ref{model}) be tested.  In Fig.\ref{fig:model} we display the
values of the cross section in the photoproduction region as a
function of the odderon coupling parameter~$c_\odd$.  Clearly, the
effect of the odderon on $\eta'$ production is much weaker than on
$\pi^0$ or $\eta$ production.  However, due to the experimental
uncertainty in the value of~$\Gamma_{\gamma\gamma}$ (cf.\
Tab.\ref{tab:mesons}), the normalization of the photon-photon cross
section is known only to $10\%$ accuracy.  Hence, in order to be
sensitive to the odderon coupling strength, it is important to
separate its contribution kinematically, as we will discuss in the
following section.

\section{Phenomenology}
\begin{figure}
\begin{center}
\includegraphics{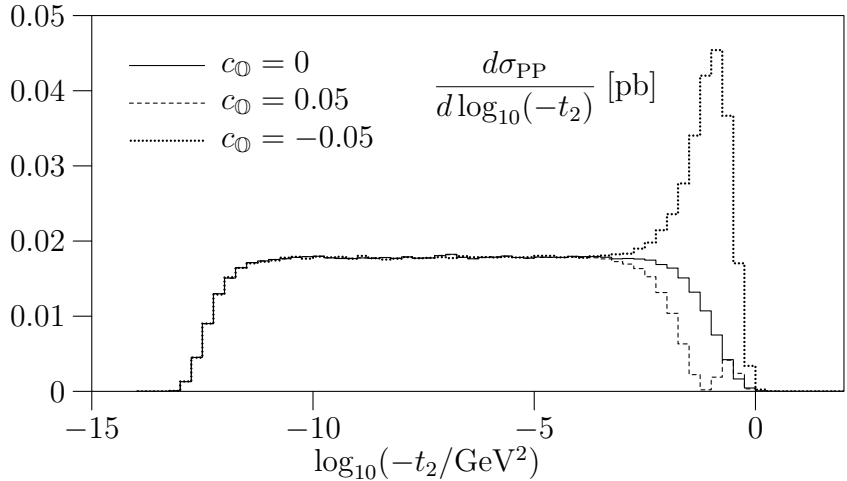}
\end{center}
\vspace*{\baselineskip}
\caption{$t_2$ distribution for pion production in the PP
region~(\ref{cuts-pp}).}
\label{fig:t2}
\end{figure}

The most characteristic feature of the hadronic interaction in
Fig.\ref{fig:process}(b) is the absence of the propagator pole in
$t_2$.  In this respect, the hadronic vertex behaves similar to a
four-particle contact term.  Its presence should therefore manifest
itself by an enhancement (or a reduction) of events with large values
of $t_2$ due to a constructive (destructive) interference of the two
diagrams in Fig.\ref{fig:process}.  This is illustrated in
Fig.\ref{fig:t2}, where we show the differential cross section with
respect to the logarithm of $t_2$ for the PP cuts~(\ref{cuts-pp}).
Whereas photon exchange results in a distribution in $\log|t_2|$ which
is constant over many decades, the hadronic contribution is
concentrated in the region between $0.01$ and $1\;{\rm GeV}^2$.
Depending on the sign~$\eta_\odd$ of the odderon coupling, there is
positive or negative interference with the photon-photon fusion
amplitude.

In diffractive scattering $t_2$ is usually not easily observable.
However, in the photoproduction region where $Q^2\approx 0$, the
transverse momentum squared~$p_\perp^2$ of the final-state meson is
approximately equal to~$-t_2$.  Thus, for negative sign, $\eta_\odd=-1$,
the odderon contribution shows up as an enhancement of the higher
$p_\perp$~values (cf.\ Fig.\ref{fig:pt}).  If $\eta_\odd$ is positive,
the odderon amplitude is of opposite sign to the photon-photon
amplitude.  This effect results in a dip in the $p_\perp$~distribution
at the value where the interference is maximal.

\begin{figure}[p]
\begin{center}
\includegraphics{pt.1}
\end{center}
\vspace*{\baselineskip}
\caption{$p_\perp$ distribution for pion production in the PP
region~(\ref{cuts-pp}).}
\label{fig:pt}

\begin{center}
\includegraphics{rap.1}
\end{center}
\vspace*{\baselineskip}
\caption{Rapidity distribution for pion production in the PP
region~(\ref{cuts-pp}).}
\label{fig:rap}
\end{figure}

The hadronic contribution is visible also in other observables: For
instance, if the PP cuts are applied, the rapidity~$\eta$ of the
final-state meson is bounded [cf.(\ref{rap})]
\begin{equation}
   \eta \leq
	\eta_\cm - \log\frac{\sqrt{s}}{m_\ps} - \log y_\mn
\end{equation}
if $t_2$ can be neglected, which holds true for the bulk of the
photon-photon cross section, but not for the odderon contribution.
Thus, the odderon affects the tail in the rapidity distribution
(Fig.\ref{fig:rap}), and a cut on~$\eta$ near $\eta_\mx$ is one
possibility for separating the odderon contribution.  A similar
consideration applies to the invariant mass $s_1$ of the
electron-meson system.

\begin{figure}[p]
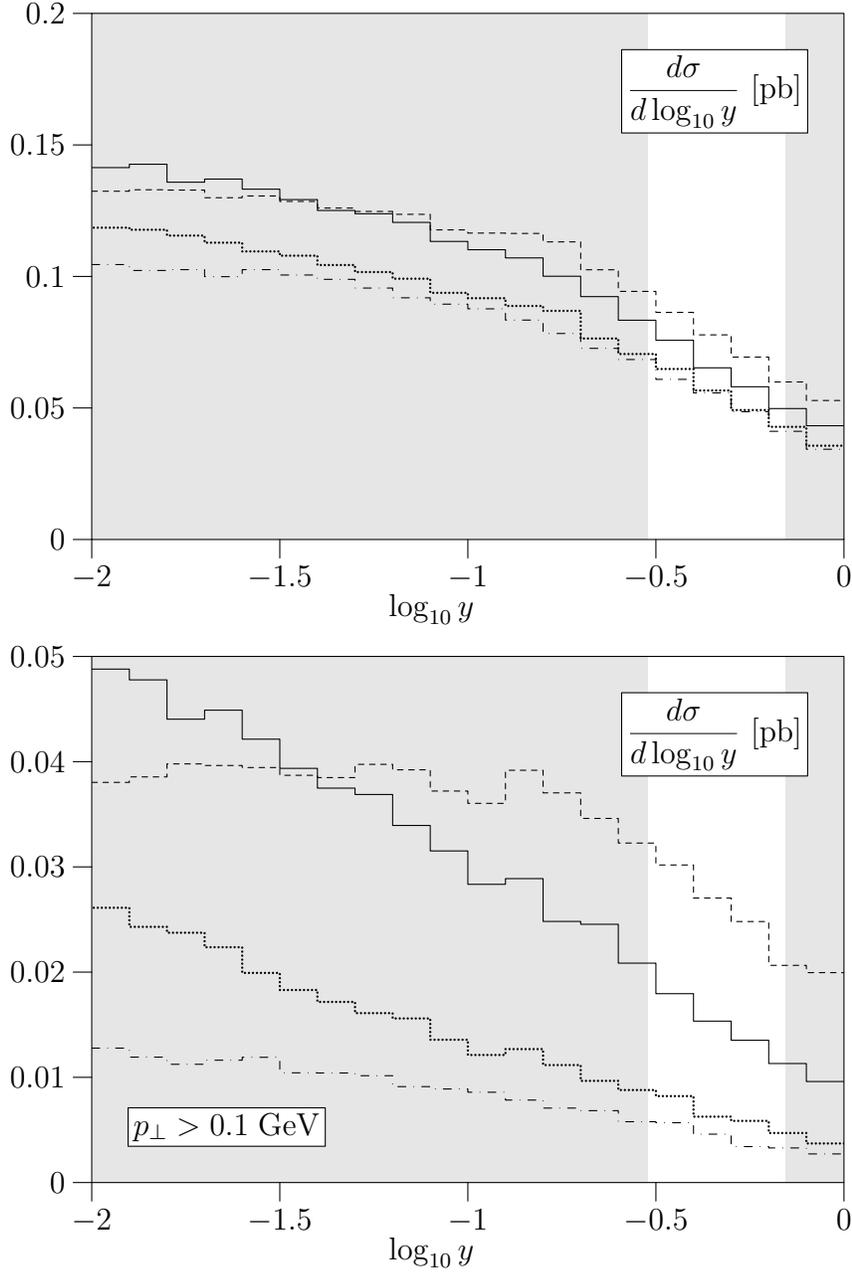

\begin{center}
\includegraphics{y.1}
\\[1.5cm]
\includegraphics{y.2}
\end{center}
\vspace*{\baselineskip}
\caption{Upper part: $\log y$ distribution for pion production in the
PP region.  Solid line: $\alpha_\odd(0)=1$, $c_\odd=-0.05$; dashed:
$\alpha_\odd(0)=1.2$, $c_\odd=-0.01$; dotted: $\alpha_\odd(0)=0.8$,
$c_\odd=-0.10$.  The other parameters are as given in~(\ref{ref}).
The lowest curve (dash-dots) corresponds to photon-photon fusion alone
($c_\odd=0$).  Lower part: The analogous distributions with a cut
$p_\perp>0.1\;{\rm GeV}$ for the produced meson applied.  The window
in $y$ selected by the cuts~(\ref{cuts-pp}) is indicated by the
unshaded band.}
\label{fig:y}
\end{figure}

These signatures are essentially independent of the particular
parameterization~(\ref{replace}) we chose for the hadronic vertex.  A
more specific test of the odderon hypothesis would be to measure the
parameters the ansatz~(\ref{replace}) depends on.

Regge phenomenology predicts a power dependence on the hadronic
subenergy:
\begin{equation}
  \frac{d\sigma}{dt_2}(t_2=0)\propto (s_2/s_0)^{\alpha_\odd(0)-1}.
\end{equation}
For the pomeron, a value of $\alpha_\pom(0)>1$ has been
observed~\cite{DL}.  This behavior may also occur for the odderon; in
addition, there are subleading contributions [\emph{e.g.}, an $\omega$
trajectory with $\alpha_\omega(0)\sim 0.5$].  Compared to photon
exchange, the hadronic contribution becomes dominant at asymptotic
energies if $\alpha_\odd(0)>1$, or ``dies out'' in the opposite case.
In order to be sensitive to the exponent $\alpha_\odd(0)$, the
hadronic subenergy $\sqrt{s_2}=W$, or, equivalently, the electron
energy loss~$y$, should be measured over several decades.  In
Fig.\ref{fig:y} we display the distribution in~$y$ for pion
photoproduction for the values $\alpha_\odd(0)=0.8$, $1$, and $1.2$.
The odderon coupling has been adjusted in each case such that the
total PP cross sections are comparable.  Without additional cuts, a
difference in the slope of the $y$~distributions is hardly detectable.
If, on the other hand, a $p_\perp$~cut of $0.1\;{\rm GeV}$ for the
produced pion is applied, the $\gamma\gamma$ background is reduced and
the shape of the three curves can easily be distinguished.  However,
in order to disentangle the measurements of $\alpha_\odd(0)$ and of
the odderon coupling~$\beta_\odd$ by this method, it is necessary that
$y$~values down to $10^{-2}$ are accessible, or, equivalently,
experiments are carried out at lower collider energies.

At HERA, for low values of $Q^2$ the hadronic subenergy is given by
$W\sim \sqrt{ys} = 300\;{\rm GeV}\times\sqrt{y}$.  Thus, non-leading
Regge terms as discussed above should be very small for $0.03\lesssim
y \leq 1$, corresponding to $50\;{\rm GeV}\lesssim W \lesssim
300\;{\rm GeV}$.  For smaller~$y$ and/or c.m.\ energies such
non-leading terms should be included in the analysis.

\begin{figure}[hbt]
\begin{center}
\includegraphics{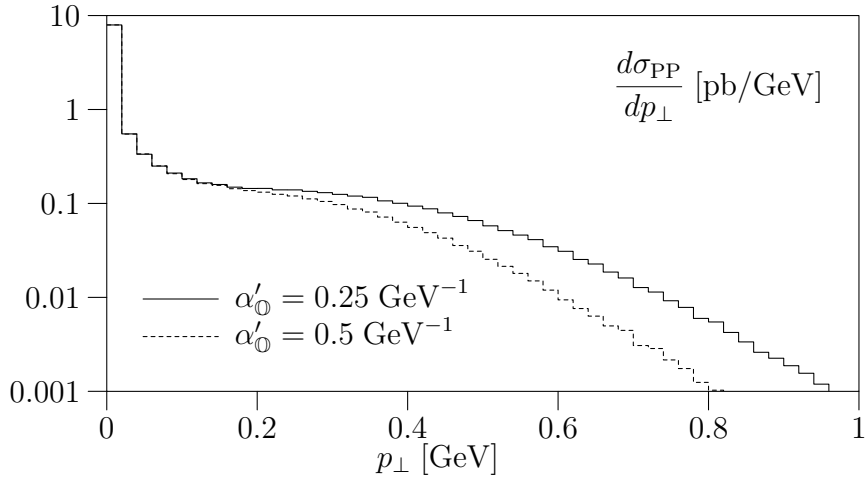}
\end{center}
\vspace*{\baselineskip}
\caption{$p_\perp$ distribution for pion production in the PP
region~(\ref{cuts-pp}).  The odderon coupling is fixed to
$c_\odd=-0.05$.}
\label{fig:pt2}
\end{figure}

With sufficient statistics, the dependence on $t_2$ can be determined
from the meson's $p_\perp$ distribution, and the odderon form factors
can be measured.  In Fig.\ref{fig:pt2} we show the influence of the
value of the parameter $\alpha_\odd'$ which controls the exponential
falloff of the odderon amplitude, Changing $\alpha_\odd'$ from
$0.25\;{\rm GeV}^{-1}$ to $0.5\;{\rm GeV}^{-1}$, results in a
reduction of events with $p_\perp>0.2\;{\rm GeV}$ if the phase of the
odderon coupling is negative.  In the opposite case ($\eta_\odd=+1$),
this change shifts the position of the dip in the $p_\perp$
distribution, the experimental observation of which, however, requires
high event rates.

\section{Transition to the hard region}

\begin{figure}
\begin{center}
\includegraphics{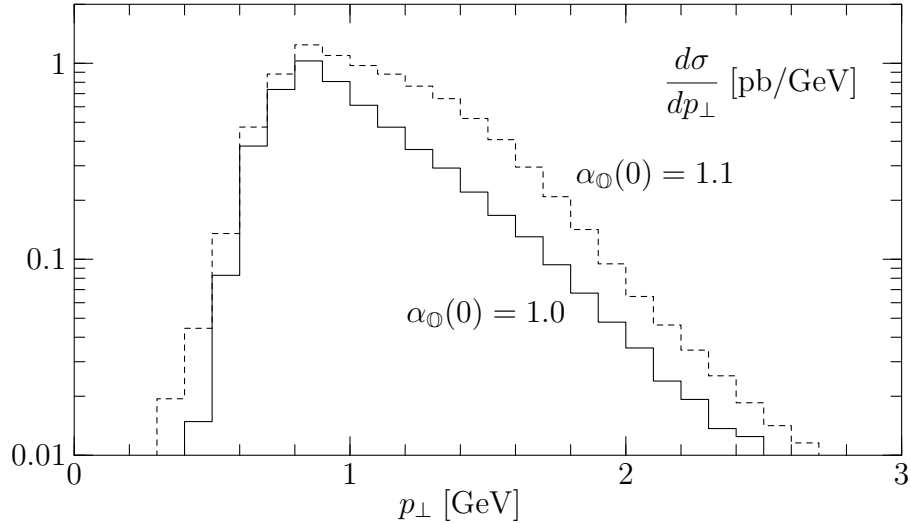}
\end{center}
\caption{$p_\perp$ distribution for $\pi^0$ production in the DIS
region~(\ref{cuts-dis}).  The two curves correspond to two different
values of the exponent~$\alpha_\odd$.  The other parameters are fixed to
their reference values~(\ref{ref}).}
\label{fig:pt-dis}
\end{figure}

In vector meson production at HERA, deviations from the universal
``soft pomeron'' behavior are observed whenever there is a hard scale
which limits the effective transverse size of the interaction
region~\cite{Gal}.  In particular, if either the initial photon has a
high virtuality or the produced meson has a large mass (\emph{e.g.},
$J/\psi,\Upsilon$) in comparison to the typical hadronic scale $\sim
1\;{\rm GeV}$, the dependence on $s$ becomes steeper.

A similar transition to the hard region should also be observable in
the processes considered in this paper.  As illustrated in
Fig.\ref{fig:pt-dis}, in the DIS region the effect of an increase in
the exponent $\alpha_\odd(0)$ shows up as a broadening of the
$p_\perp$~distribution, since an enhancement of the odderon
contribution with respect to the $\gamma\gamma$ amplitude allows for
higher values of the momentum transfer~$t_2$.  If the $p_\perp$
distribution in the DIS region is compared to the corresponding
distribution in photoproduction, it is possible in principle to
isolate the effect of a hard scale in the interaction.  Furthermore,
the transition form factor of the $\eta_c$ meson is expected to
decrease more slowly than the form factors of the light
mesons~\cite{FK}.  For this reason, the production cross section for
the $\eta_c$~meson in the DIS region is of comparable size to the
cross section of light pseudoscalar mesons (cf.\ Tab.\ref{tab:tot}),
and the observed $\eta_c$~production rate can serve as another probe
of a possible hard odderon\footnote{A perturbative estimate for
$\eta_c$~production via odderon exchange can be found
in~\cite{CKMS}.}, if total cross sections of magnitude $1\;{\rm pb}$
or less are within the experimental reach.  [Note that an increase of
the accessible $y$~range would enhance the observable rates
considerably, similar to the photoproduction case as illustrated in
Fig.\ref{fig:y}.]

\section{Conclusions}
In this article we have discussed in detail exclusive pseudoscalar
meson production in $e^\pm p$~scattering at HERA energies.  For the
contribution from $\gamma\gamma$-fusion we gave numerical values
obtained from two versions of the equivalent photon approximation: (i)
applied to both the photon from the electron line and the proton line
(DEPA) and (ii) applied to the photon from the electron line only
(EPA).  The full calculation showed that DEPA (EPA) give accuracies of
order 20\% ($<5\%$) for total cross sections without cuts and with
photoproduction cuts~(\ref{cuts-pp}).  For the deep-inelastic
scattering cuts~(\ref{cuts-dis}) only the full calculation is
reliable\footnote{The Monte Carlo programs used for the
calculations are available from the authors.}.

In Sec.\ref{sec:odderon} we introduced a simple ansatz for a possible
odderon exchange contribution to exclusive pseudoscalar meson
production.  We showed how this odderon affects total cross sections
and differential distributions.  Interference effects are large in the
parameter region of interest and have been taken into account.
Particularly promising as signals for the odderon are the $t_2$ and
$p_\perp$ distributions (Figs.\ref{fig:t2}--\ref{fig:pt}).  A larger
acceptance in the variable $y$~(\ref{HERA}) than assumed
in~(\ref{cuts-pp},\ref{cuts-dis}) would increase the cross sections
considerably (Fig.\ref{fig:y}).  This would lead to a greatly
increased potential for finding the odderon and determining its
parameters.

In the HERA kinematical situation, events with $Q^2\lesssim 10\;{\rm
GeV}^2$ and $0.03\lesssim y \leq 1$ correspond to a hadronic
energy~$W$ between $50$ and $300\;{\rm GeV}$.  This should be large
enough for non-leading Regge terms to be suppressed.  However, it is
straightforward to include such terms in the theoretical formulae for
a detailed analysis of experimental data.

The cross sections we have calculated for pseudoscalar meson
production at HERA are between $0.5$ and $1800\;{\rm pb}$, depending
on the produced meson and on the particular cuts imposed (cf.\
Tab.\ref{tab:tot} and Fig.\ref{fig:model}).  With an integrated
luminosity of $30\;{\rm pb}^{-1}$ per year such processes should
definitely be observable and give valuable insight in the mechanisms
of diffraction in QCD.

\enlargethispage{\baselineskip}
\subsection*{Acknowledgements}
We are grateful to T.~Berndt, H.-G.~Dosch, P.V.~Landshoff, K.~Meier,
T.~Ohl, M.~R\"uter, R.~Ruskov, and S.~Tapp\-rogge for fruitful
discussions and valuable comments.  We also thank H.-G.~Dosch and
P.V.~Landshoff for a critical reading of the manuscript.

\end{fmffile}


\baselineskip15pt
\begin{thebibliography}{19}
\bibitem{LN}
	F.E.~Low,
	Phys.\ Rev.\ {\bf D12} (1975) 163;
	S.~Nussinov,
	Phys.\ Rev.\ Lett.\ {\bf 34} (1975) 1286.
\bibitem{CW}
	H.~Cheng and T.T.~Wu,
	\emph{Expanding Protons},
	MIT Press, Cambridge, Mass.\ (1987)
	and references therein.
\bibitem{GS}
	J.F.~Gunion and D.E.~Soper,
	Phys.\ Rev.\ {\bf D15} (1977) 2617.
\bibitem{BFKL}
	E.A.~Kuraev, L.N.~Lipatov, and V.S.~Fadin,
	Sov.\ Phys.\ JETP {\bf 44} (1976) 443;
	L.N.~Lipatov, 
	Sov.\ J.\ Nucl.\ Phys.\ {\bf 23} (1976) 338;
	Y.Y.~Balitskij and L.N.~Lipatov,
	\emph{ibid.} {\bf 28} (1978) 822.
\bibitem{LNm}
	P.V.~Landshoff and O.~Nachtmann,
	Z.\ Phys.\ {\bf C35} (1987) 405.
\bibitem{Nm}
	O.~Nachtmann,
	Ann.\ Phys.\ (N.Y.) {\bf 209} (1991) 436.
\bibitem{DFK}
	H.G.~Dosch, E.~Ferreira, and A.~Kr\"amer,
	Phys.\ Rev.\ {\bf D50} (1994) 1992.
\bibitem{NmS}
	O.~Nachtmann,
	\emph{High Energy Collisions and Nonperturbative QCD},
	in: 
	\emph{Perturbative and Nonperturbative Aspects of 
	Quantum Field Theory},
	H.~Latal, W.~Schweiger (eds.),
	Springer Verlag (1997).
\bibitem{Col}
	P.D.B.~Collins,
	\emph{An Introduction to Regge Theory},
	Cambridge University Press (1977);
	L.~Caneschi (ed.),
	\emph{Regge Theory of low $p_T$ hadronic interaction},
	North Holland, Amsterdam (1989).
\bibitem{DL}
	A.~Donnachie and P.V.~Landshoff,
	Nucl.\ Phys.\ {\bf B244} (1984) 322;
	\emph{ibid.} {\bf B267} (1986) 690;
	Phys.\ Lett.\ {\bf B185} (1987) 403.
\bibitem{PDG}
	R.M.~Barnett \emph{et al.} (Particle Data Group),
	Phys.\ Rev.\ {\bf D54} (1996) 1.
\bibitem{VH}
	L.~Van Hove, 
	Phys.\ Let.\ {\bf B24} (1967) 183.
\bibitem{MP}
	C.J.~Morningstar and M.~Peardon,
	Phys.\ Rev.\ {\bf D56} (1997) 4043;
	M.~Teper,
	Preprint OUTP--97--66P, hep-ph/9711299,
	to appear in: 
	\emph{Proceedings of the Europhysics HEP Conference},
	Jerusalem, August 1997.
\bibitem{Lan}
	P.V.~Landshoff,
	\emph{The two pomerons},
	in:
	\emph{Proceedings of the Summer School on Hadronic Aspects
	of Collider Physics, Zuoz, 1994},
	M.P.~Locher (ed.), PSI, Villigen (1994).
\bibitem{Lip}
	L.N.~Lipatov,
	\emph{Pomeron in Quantum Chromodynamics},
	in: \emph{Perturbative Quantum Chromodynamics},
	A.H.~Mueller (ed.),
	World Scientific, Singapore (1989).
\bibitem{FR}
	J.R.~Forshaw and D.A.~Ross,
	\emph{Quantum chromodynamics and the pomeron},
	Cambridge University Press (1997).
\bibitem{JLN}
	L.~Lukaszuk and B,~Nicolescu,
	Nuov.\ Cim.\ Lett.\ {\bf 8} (1973) 405;
	D.~Joynson, E.~Leader, C.~Lopez, and B.~Nicolescu,
	Nuov.\ Cim.\ {\bf 30A} (1975) 345.
\bibitem{BGN}
	D.~Bernard, P.~Gauron, and B.~Nicolescu,
	Phys.\ Lett.\ {\bf B199} (1987) 125;
	P.~Gauron, E.~Leader, and B.~Nicolescu,
	Phys.\ Rev.\ Lett.\ {\bf 54} (1985) 2656;
	\emph{ibid.} {\bf 55} (1985) 639.
\bibitem{Lea}
	E.~Leader, 
	Phys.\ Lett.\ {\bf B253} (1991) 457.
\bibitem{CDGJP}
	R.J.M.~Corolan, P.~Desgrolard, M.~Giffon, L.L.~Jenkovszky,
	and E.~Predazzi,
	Z.\ Phys.\ {\bf C58} (1993) 109.
\bibitem{Lip90}
	L.N.~Lipatov,
	Preprint DESY 90--060 (unpublished);
	P.~Gauron, L.~Lipatov, and B.~Nicolescu,
	Phys.\ Lett.\ {\bf B304} (1993) 334.
\bibitem{AB}
	N.~Armesto and M.A.~Braun,
	Preprint DESY-97-150, hep-ph/9708296.
\bibitem{Str}
	B.V.~Struminskii,
	Phys.\ Atom.\ Nucl.\ {\bf 57} (1994) 1398.
\bibitem{JW}
	R.A.~Janik and J.~Wosiek,
	Phys.\ Rev.\ Lett.\ {\bf 79} (1997) 2935.
\bibitem{DLo}
	A.~Donnachie and P.V.~Landshoff,
	Nucl.\ Phys.\ {\bf B348} (1991) 297.
\bibitem{Gin}
	I.F.~Ginzburg,
	JETP Lett.\ {\bf 59} (1994) 605.
\bibitem{DR}
	H.G.~Dosch and M.~Rueter,
	Phys.\ Lett.\ {\bf B380} (1996) 177.
\bibitem{SMN}
	A.~Sch\"afer, L.~Mankiewicz, and O.~Nachtmann,
	in: \emph{Proceedings of the 1991 DESY/HERA Workshop} 
	(W.~Buchm\"uller and G.~Ingelman, eds.).
\bibitem{ST}
	S.~Tapprogge, Ph.D.~Thesis, 
	Report HD--IHEP 96--19, and private communication.
\bibitem{BL81}
	S.J.~Brodsky and G.P.~Lepage,
	Phys.\ Rev.\ {\bf D24} (1981) 1808.
\bibitem{AMN}
	V.V.~Anisovich, D.I.~Melikhov, and V.A.~Nikonov,
	Phys.\ Rev.\ {\bf D55} (1997) 2918.
\bibitem{RR}
	A.V.~Radyushkin and R.~Ruskov,
	Phys.\ Lett.\ {\bf B374} (1996) 173;
	Nucl.\ Phys.\ {\bf B481} (1996) 625.
\bibitem{Exp}
	H.-J.~Behrend \emph{et al.} (CELLO Collaboration),
	Z.~Phys.\ {\bf C49} (1991) 401;
	H.~Aihara \emph{et al.} (TPC/2$\gamma$ Collaboration),
	Phys.\ Rev.\ Lett.\ {\bf 64} (1990) 172;
	V.~Savinov (CLEO Collaboration),
	in: \emph{Proceedings of the 10th Workshop on Photon-Photon
	Collisions (PHOTON '95)}, Sheffield, England, April 8-13 1995.
\bibitem{FK}
	T.~Feldmann and P.~Kroll,
	Preprint WU-B-97-22, hep-ph/9709203.
\bibitem{KWZ}
	G.~K\"opp, T.F.~Walsh, and P.M.~Zerwas,
	Nucl.\ Phys.\ {\bf B70} (1974) 461;
	S.J.~Brodsky and G.P.~Lepage,
	Phys.\ Rev.\ {\bf D22} (1980) 2157.
\bibitem{BGMS}
	V.M.~Budnev, I.F.~Ginzburg, G.V.~Meledin, and V.G.~Serbo,
	Phys.\ Rep.\ {\bf 15} (1975) 181.
\bibitem{pFF}
	F.~Borkowski \emph{et al.},
	Nucl.\ Phys.\ {\bf B93} (1975) 461;
	P.E.~Bosted \emph{et al.},
	Phys.\ Rev.\ Lett.\ {\bf 68} (1992) 3841.
\bibitem{CompHEP}
	E.E.~Boos, M.N.~Dubinin, V.A.~Ilyin, 
	A.E.~Pukhov, and V.I.~Savrin,
	Report SNUTP--94--116, hep-ph/9503280 (unpublished);
	P.A.Baikov \emph{et al.}, 
	in: \emph{Proceedings of the Workshop QFTHEP-96}, 
	eds.\ B.~Levtchenko and V.~Savrin, 
	(Moscow 1996), hep-ph/9701412.
\bibitem{Sch}
	G.~Schuler,
	Preprint CERN--TH 96/313, hep-ph/9611249;
	Preprint CERN--TH 97/265, hep-ph/9710506.
\bibitem{UA4}
	C.~Augier \emph{et al.} (UA 4/2 Collab.),
	Phys.\ Lett.\ {\bf B316} (1993) 448;
	N.A.~Amos \emph{et al.} (E--710 Collab.),
	Phys.\ Rev.\ Lett.\ {\bf 68} (1992) 2433.
\bibitem{Gal}
	E.~Gallo,
	to appear in:
	\emph{Proceedings of the XVIII International Symposium on
	Lepton-Photon interactions},
	Hamburg, July 1997.
\bibitem{CKMS}
	J.~Czy\.{z}ewski, J.~Kwieci\'{n}ski, L.~Motyka, and M.~Sazikowski,
	Phys.\ Lett.\ {\bf B398} (1997) 400,
	rev.\ in: hep-ph/9611225.
\end{thebibliography}
\end{document}